\documentclass[12pt]{article}
\usepackage{amssymb}
\usepackage{amsthm}
\usepackage{amsmath}
\usepackage{graphicx}

\graphicspath{{./Figs/}}
\usepackage{lmodern}
\usepackage{graphics}
\usepackage{float}

\newtheorem{teo}{Theorem}
\newtheorem{lem}{Lemma}

\newcommand{\pa}{\partial}

\newcommand{\vp}{\varphi}
\newcommand{\ve}{\varepsilon}
\newcommand{\om}{\omega}
\newcommand{\be}{\begin{equation}}
\newcommand{\ee}{\end{equation}}

\newcommand{\ipd}{\stackrel{\normalfont\text{def}}{=}}
\newcommand{\const}{\operatorname{const}}

\usepackage[usenames]{color}
\begin{document}
\allowdisplaybreaks
\title{Solitary wave solutions of a generalization of the mKdV equation}
\author{J.~Noyola Rodriguez
\thanks{Universidad Aut\'onoma de Guerrero, Carlos E. Adame 54, 39650 Acapulco de Ju\'arez, Guerrero, Mexico,\
20264@uagro.mx}\and  G.~Omel'yanov*\thanks{
Corresponding author, Universidad de Sonora, Rosales y Encinas, 83000 Hermosillo, Sonora, Mexico,\ omel@mat.uson.mx} }
\date{}
\maketitle
\begin{abstract}
We consider  a generalization of the mKdV equation, which contains dissipation terms similar to those contained in both the Benjamin-Bona-Mahoney equation and the famous Camassa-Holm  and Degasperis-Procesi equations. Our  objective is  the construction of  classical (solitons) and non-classical (peakons and cuspons) solitary wave solutions of this equation.
\end{abstract}
\emph{Key words}:  general mKdV model, Camassa-Holm equation, Degasperis-Procesi equation, soliton,
peakon, cuspon

\emph{2010 Mathematics Subject Classification}: 35Q35, 35Q53, 35D30

\section{Introduction}
 We consider a generalization of the modified Korteweg-de Vries (gmKdV) equation
\begin{align}
&\frac{\partial }{\partial t}\left\{u-\alpha^2\ve^2\frac{\pa^2 u}{\pa x^2}\right\}\label{1}\\
&+\frac{\pa}{\pa x}\left\{c_0u+c_1u^3
-c_2\ve^2\Big(\frac{\pa u}{\pa x}\Big)^2+\ve^2\big(\gamma-c_3u\big)\frac{\pa^2 u}{\pa x^2}\right\}=0, \; x \in \mathbb{R}^1, \; t > 0.\notag
\end{align}
which describes  unidirectional propagations  of shallow water waves.  Here $\alpha$, $c_0,\dots,c_3$, $\gamma$ are  real parameters  and  $\varepsilon$ characterizes the level of dispersion.  The constants $\alpha\geq0$ and $\gamma\geq0$ are associated with different characters of the linear dispersion manifestation, whereas the  terms with $c_2\geq0$ and $c_3\geq0$ can be treated as representations of ``nonlinear dispersion".  In the Green-Naghdi approximation  $c_2+c_3>0$ \cite{GN}.

  It is obvious that for $\alpha=c_2=c_3=0$ equation (\ref{1}) coincides with the mKdV equation. The main feature of the  inclusion of the ``nonlinear dispersion" terms in the gmKdV model is the description of a  fundamental phenomenon in the theory of water waves: the appearance of the breaking effect.
 Such a mechanism has been studied in detail for the famous Camassa-Holm (CH) equation (with $u^2$ instead of $u^3$, $c_2=c_3/2$,  and $\gamma=0$; see \cite{CH}-\cite{ConLan}); and for the Degasperis-Procesi (DP) equation (with $u^2$ instead of $u^3$, $c_2=c_3$,  and $c_0=\gamma=0$; see  \cite{ConLan}-\cite{ELY}). The same should be true for the equation (\ref{1}) in view of the  balance law
\be
\frac{d}{dt}\int_{-\infty}^\infty\{u^2+\alpha^2(\ve u_x)^2\}dx+\ve^{-1}(2c_2-c_3)\int_{-\infty}^\infty(\ve u_x)^3dx=0,\label{1c}
\ee
which makes (\ref{1}) related to the ``general Degasperis-Procesi" (gDP) model (with $u^2$ instead of $u^3$, see \cite{NoyOm}-\cite{NoyOm1}).

Furthermore, it is well known that the CH and DP (as well as the KdV) equations are completely integrable and admit ``long-living" solutions: solitons and  continuous solitary waves called peakons (the first derivative is bounded) and cuspons (the first derivative is unbounded). It is well known also that these waves collide in the ``elastic" manner (like KdV solitons, see e.g. \cite{CH, CH1, BSSz},
\cite{ DegProc}-\cite{Lund}, and \cite{Len}-\cite{ZQQ}). Similar solitary wave solutions were also constructed for the gDP equation \cite{NoyOm, Om}. In addition,  for non-integrable cases, it was proved for $\ve<<1$ that gDP solitons collide ``almost elastically": they pass through each other, but with the appearance of a small
oscillating tail, the so-called ``effect of radiation" \cite{Om1, NoyOm1}.

The main object of the present paper is  to construct solitary wave solutions for gmKdV equation (\ref{1}). It turned out that the correspondence between the equations gmKdV and gDP is more unexpected than the correspondence between the equations mKdV and KdV. Firstly, as it proved,  in the gmKdV model there are two different mechanisms for the formation of solitons and anti solitons, however, in contrast to gDP and KdV equations, both waves move with positive velocities. Further, if we neglect the Benjamin-Bona-Mahoney effect, assuming that $\alpha = 0$, then a very strange cuspon wave formation is detected: by setting the initial condition, we should determine the cuspon amplitude  as an only one possible value, but the initial wave profile can be set almost arbitrarily, and as a result, there appears  a family of waves of the same amplitude but with almost arbitrary propagation speeds.
Let us recall that the standard process of a self-similar wave construction is as follows: by setting an initial wave amplitude,  we uniquely  determine both the wave profile and wave velocity.

In order to construct soliton-type solutions of gmKdV equation we use the approach developed in \cite{NoyOm}.
Concerning the weak solution construction, we use an approach based on the algebraic point of view. Indeed, non-classical traveling waves $u=u(x-Vt)$ of (\ref{1}) should be distributions such that $(u(\eta)'_\eta)^2\in \mathfrak{D}'(\mathbb{R}^1)$, in other words $u(\eta)$ and $u(\eta)'_\eta$ should belong to a subalgebra in $\mathfrak{D}'(\mathbb{R}^1)$. We use two of them.  The first one has the generators  $\{\textbf{1}, H(\eta)\}$, where $\textbf{1}$ denotes the space of smooth functions and $H(\eta)$ is the Heaviside function: $H(\eta)=0$ for $\eta<0$, and $H(\eta)=1$ for $\eta>0$. The Heaviside function is associated with the sequence
\be
\dots,\quad \eta_+,\quad H(\eta),\quad \delta(\eta),\quad \delta'(\eta),\dots,\label{3}
\ee
where $\eta_+=\eta H(\eta)$; $\delta(\eta)$ and $\delta'(\eta),\dots$ are the Dirac delta-function and its derivatives. This subalgebra allows us to construct peakon-type solutions.

The second subalgebra has the generators  $\{\textbf{1}, \eta_+^\lambda\}$, where $\lambda\in (0,1)$ (see e.g. \cite{GelfShil}). Respectively, the distribution $\eta_+^{\lambda}$ is associated with the sequence
\be
\dots,\quad \eta_+^{\lambda+2},\quad \eta_+^{\lambda+1},\quad \eta_+^{\lambda},\quad \eta_+^{\lambda-1},\quad \eta_+^{\lambda-2},\dots,\label{4}
\ee
and it allows us to construct cuspon-type solutions.

 In what follows we assume
\be\label{5}
\gamma\geq0,\quad \alpha\geq0, \quad \gamma+\alpha>0,\quad c_0\geq0,\quad c_k>0,\quad k=1,2,3,
\ee
 and treat $\ve\neq0$ as a fixed  parameter.

The paper contents is the following: Section 2 is devoted to solitons, in Subsection 3.1 we present the construction of peakons and obtain explicit formulas for such waves. Cuspons are considered in Subsection 3.2.  In addition, in each section we describe the procedure for numerical calculation of the corresponding wave. In Conclusion,  we summarize all the results found for solitary wave solutions of (\ref{1}) and present a list of open problems for this equation.
\section{Smooth solitary waves}

The soliton construction seems to be quite traditional. We set the ansatz
\begin{equation}\label{205}
u=A\om\big(\beta(x-Vt)/\varepsilon,A\big),
\end{equation}
where $\om(\eta,A)$ is a smooth  function such that
\begin{align}
&\om(-\eta,A)=\om(\eta,A),\quad\om(\eta,A)\to0\quad\text{as}\quad \eta\to\pm\infty,\label{206}\\
&\om(0,A)=1,\label{207}
\end{align}
the amplitude $A$ and the scale $\beta$ are free parameters, and the velocity
 $V=V(A)\neq 0$ should be determined.

Let
\be
\gamma+\alpha^2V>0. \label{206a}
\ee
Then, substituting  (\ref{205}) into Eq.(\ref{1}), integrating, and using the second assumption in (\ref{206}), we obtain the following version of the inverse scattering problem:

\textit{Determine the velocity} $V$ \textit{such that the equation}
\begin{align}
\Big\{1-&\frac{c_3A }{\gamma+\alpha^2V}\om\Big\}\frac{d^2 \om}{d \eta^2}=\frac{c_2A }{\gamma+\alpha^2V}\left(\frac{d \om}{d \eta}\right)^2\notag\\
&+\frac{1}{\beta^2(\gamma+\alpha^2V)}\big((V-c_0)\om-c_1A^2\om^3\big),\quad \eta\in \mathbb{R}^1,\label{209}
\end{align}
\textit{admits a nontrivial smooth solution with the properties} (\ref{206}) and (\ref{207}).

To simplify formulas we  choose  the scale
\begin{equation}\label{205a}
\beta=\sqrt{c_1(\gamma+\alpha^2V)/rc_3^2},\quad\text{where}\quad r=c_3/(c_2+c_3),
\end{equation}
and define the notation
\begin{equation}
 W=p\om,\; p=\frac{c_3A}{\gamma+\alpha^2V},\quad
 q=\frac{c_3^2(V-c_0)}{c_1(\gamma+\alpha^2V)^2}.\label{2012}
\end{equation}
Then we transform  the equation (\ref{209}) to the following form
\begin{equation}\label{2013}
(1-W)\frac{d^2 W}{d\eta^2}=\frac{1-r}{r}\left(\frac{d W}{d\eta}\right)^2+r(q\,W-W^3),\quad \eta\in \mathbb{R}^1.
\end{equation}
The terms $WW''$ and $(W')^2$ prevent integration of (\ref{2013}) in a standard way. To avoid this obstacle, we use substitution \cite{NoyOm}
\begin{equation}\label{2014}
 W(\eta)=1-g(\eta)^r,
\end{equation}
which allows us to eliminate the first derivative from the model equation (\ref{2013}). Consequently, after the integration we obtain the first order ODE
\be
\Big(\frac{d g}{d\eta}\Big)^2=F(g,q),\quad \eta\in{\mathbb{R}}^1,\label{2018}
 \ee
where
\be
F(g,q)=\frac{1-q}{1-r}g^{2-2r}-2\frac{3-q}{2-r}g^{2-r}+3g^2-\frac{2}{2+r}g^{2+r}-C(q).\label{2019}
\ee
To satisfy the second assumption in (\ref{206}), let us choose the constant of integration $C(q)$ setting $F|_{g=1}=0$. Then
\be
C(q)=\frac{r}{(1-r)(4-r^2)}\{3r^2-q(2+r)\}.\label{2020}
\ee
Furthermore, simple calculations imply the equality
\be
\frac{dF}{dg}=2g^{1-r}(g^{-r}-1)(g^{r}-g_0^*)(g^{r}-g_1^*),\label{20191}
\ee
where
\be
g_0^*=1-\sqrt q,\quad g_1^*=1+\sqrt q.\label{20192}
\ee
Now we assume the inequality
\be
q>0,\label{20193}
\ee
which guarantees both $g_k^*\in \mathbb{R}$  and  the fulfillment of condition $V>c_0\geq0$, which ensures that the assumption (\ref{206a}) is satisfied. Simple calculations imply
\be
\frac{d^2F}{dg^2}\big|_{g=1}=2rq>0,\quad \frac{d^2F}{dg^2}\big|_{g^r=g_k^*}=-4r(g_k^*)^{-1}q<0,\quad k=0,1.\label{20194}
\ee
Since $F\to-\infty$ as $g\to\infty$, and $F|_{g=0}<0$ for $C>0$, we obtain that there exist three zero points, $g=g_0$, $g=1$, and $g=g_1$, of the right-hand side $F(g,q)$, see Fig.1.
\begin{figure}[H]\label{fig1}
\centering
\includegraphics[width=8cm]{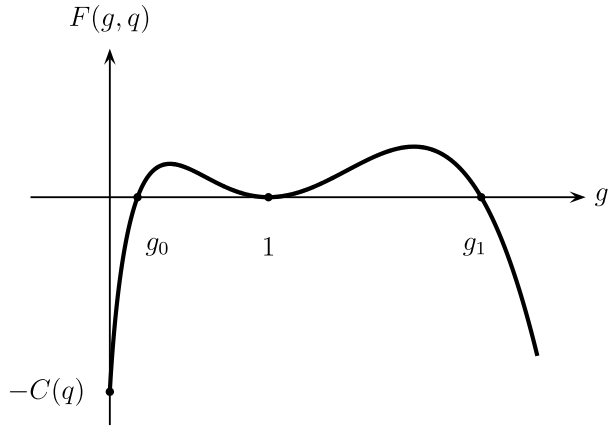}
\caption{Right-hand side of the equation (\ref{2018}) in the case $r=1/2$, $q\approx0.148$. Here $g_0\approx0.175$, $g_1\approx2.455$, and $C(q)\approx1.964$.}
%\label{f1}
\end{figure}
Recall now that a solution of the equation
\be
\frac{d g}{d\eta}=\sqrt{F(g,q)},\quad \eta\in(0,\infty),\label{2018aa}
 \ee
can be continued onto left semi-axis  in a  smooth even manner if and only if all odd derivatives $g^{2k+1}|_{\eta=0}$ are  zero. For the equation (\ref{2018aa}) this means that $g(0)$ should be a zero point of $F$.
Thus, in contrast with the standard situation (like equations KdV, mKdV, gDP and others, see e.g. \cite{NoyOm}), it is possible now to construct two different solutions, for $g\in(g_0,1)$ and $g\in(1,g_1)$. We are considering these options separately.
\subsection{Problem A, solitons for $g\in(g_0,1)$, $g_0>0$.}
Suppose
\be
C>0,\label{20194a}
\ee
then the condition
\be
q<\frac{3r^2}{2+r}\label{2017}
\ee
appears. For $r\in(0,1)$ we have $3r^2<(2+r)$, so that $q<1$; and automatically $g_0\in(0,g_0^*)$ for $q>0$. Thus,
 we can pass from the inverse scattering problem (\ref{209}) to   the equation (\ref{2018aa}) supplemented by the initial condition
\be
 g|_{\eta=0}=g_0.\label{2018a}
 \ee
In view of  denotation (\ref{2012}), (\ref{2014}), the assumption (\ref{207}) implies
 \be
1-g_0^r \ipd p_0=\frac{c_3A}{\gamma+\alpha^2V}.\label{118}
\ee
 Obviously, $F(g,\cdot)\in \mathcal{C}(\mathbb{R}_+)$, thus the solution of the problem (\ref{2018aa}), (\ref{2018a})  exists for $\eta\geq0$ and any  $q=\const\in(0,1)$, however,  it is unique for $\eta\geq\const>0$ only since $F(g,\cdot)$ doesn't satisfy the Lipschitz condition for $g|_{\eta=0}=g_0$. Indeed, the problem (\ref{2018aa}), (\ref{2018a}) has two solutions: $g\equiv g_0$ and an increasing function.

Note now that in view of (\ref{2014}), (\ref{2018a}), and (\ref{118})
\be
\om|_{\eta=0}=1,\quad \om'|_{\eta=0}=0, \quad \om''|_{\eta=0}=-\frac{r}{2p_0}g^{r-1}F'(g,q)|_{g=g_0}<0,\label{118a}
\ee
where the prime denotes the derivative with respect to $\eta$. Obviously, the function $\om(\eta)$ admits the smooth even continuation on the negative half-axis.

Note next that for $g=1-z$, $0<z<<1$, the equation (\ref{2018aa}) yields
\be
z'=-z\sqrt{rq}.\label{118b}
\ee
Thus, for  $\eta>>1$ we obtain
\be
g(\eta)\sim 1-\exp(-\sqrt{rq}\eta), \quad \om(\eta)\sim \exp(-\sqrt{rq}\eta),\label{118c}
\ee
which implies that $\om(\eta)$ satisfies the assumptions (\ref{206}).

Now it remains only to analyze the restrictions (\ref{20193}), (\ref{2017}). Let
\be
\alpha>0.\label{118cc}
\ee
Then the equality (\ref{118}) allows us to obtain the relation between the velocity and the wave amplitude:
\be
V=\frac{1}{\alpha^2}\Big\{\frac{c_3}{p_0}A-\gamma\Big\}.\label{118d}
\ee
Consequently, this and the last equality in (\ref{2012}) imply the following representation of the coefficient $q$:
\be
q\ipd q(g_0,A)=\frac{p_0}{c_1\alpha^2A^2}\big\{c_3A-p_0\gamma_\alpha\big\},\label{118e}
\ee
where
\be
\gamma_\alpha=\gamma+c_0\alpha^2.\label{118f}
\ee
Recall that our choice of the initial datum in (\ref{2018a}) assumes that $p_0>0$. This and  (\ref{20193}), (\ref{118e})
imply the condition $A>A_0^*$, where
\be
A_0^*=p_0\frac{\gamma_\alpha}{c_3}.\label{118j}
\ee
Next, the assumption (\ref{2017}) for $q$ of the form (\ref{118e}) is equivalent to the inequality
\be
\xi A^2-p_0c_3A+p_0^2\gamma_\alpha>0,\quad \xi=3r^2\alpha^2c_1/(2+r).\label{118x}
\ee
Let
\be
c_3^2>4\xi\gamma_\alpha.\label{118z}
\ee
Then (\ref{118x}) requires:  $A<A_0^-$ or $A>A_0^+$, where
\be
 A_0^\pm=\frac{p_0}{2\xi}\big(c_3\pm\sqrt{c_3^2-4\xi\gamma_\alpha}\big),\quad\label{118h}
\ee
Obviously, $q(g_0,A_0^*)=0$, $q'|_{A=A_0^*}>0$, whereas $q(g_0,A_0^\pm)>0$. Thus, $A_0^*<A_0^\pm$
 and we obtain the restriction for the case (\ref{118cc}), (\ref{118z})
\be
A\in(A_0^*,A_0^-)\bigcup(A_0^+,\infty).\label{118g}
\ee
It is clear that for
\be
c_3^2=4\xi\gamma_\alpha,\label{118zz}
\ee
instead of (\ref{118g}) we get
\be
A>A_0^*,\quad A\neq \overline{A}_0^\pm,\quad\text{where}\quad\overline{A}_0^\pm=p_0c_3/2\xi,\label{118gg}
\ee
whereas for
\be
c_3^2<4\xi\gamma_\alpha,\label{118zzz}
\ee
we assume only
\be
A>A_0^*.\label{118ggg}
\ee
Finally, we obtain that the amplitude $A$ defines in the case (\ref{118cc}) both the velocity $V$ (\ref{118d}) and the coefficient $q=q(g_0, A)$ (\ref{118e}). Thus, to complete the statement of the problem (\ref{2018aa}), (\ref{2018a}) it remains  to determinate $g_0=g_0(A)$ as the root of the equation
\be
F(g_0,q(g_0,A))=0\quad \text{with}\quad g_0\in(0,1).\label{118jj}
\ee

Assume now
\be
\alpha=0.\label{118ccc}
\ee
Then the equality (\ref{118}) uniquely defines the root $g_0=\bar{g_0}(A)$
\be
\bar{g_0}(A)=(1-c_3A/\gamma)^{1/r}.\label{118cccc}
\ee
Consequently, instead of (\ref{118jj}) we obtain the inverse problem: find a coefficient $q=q(A)$ such that the $\bar{g_0}(A)$ will be the root of $F$,
\be
F(\bar{g_0}(A),q)=0.\label{118jjj}
\ee
Obviously, to satisfy the condition $\bar{g_0}\in(0,1)$ we should assume
\be
A\in(0,\bar{A_0}^*), \quad \bar{A_0}^*=\gamma/c_3.\label{118gggg}
\ee
It is clear also that by determining the coefficient $q=q(A)$ we get both the velocity
\be
V=c_0+\frac{c_1\gamma^2}{c_3^2}q\big(\bar{g_0}(A)\big),\label{118dd}
\ee
and the solitary wave profile.

We come to the following statement
\begin{lem}
Under the assumptions (\ref{5}), (\ref{118cc}) we assume the fulfilment of the restrictions (\ref{118z}), (\ref{118g}); or (\ref{118zz}), (\ref{118gg}); or (\ref{118zzz}), (\ref{118ggg}).
 Under the assumptions (\ref{5}), (\ref{118ccc}) we assume the fulfilment of the condition (\ref{118gggg}). Then the Cauchy problem (\ref{2018aa}), (\ref{2018a})  determines  the soliton solution (\ref{205}) with the velocity $V=V(A)>c_0$ defined in (\ref{118d}) in the case (\ref{118cc}) and in (\ref{118dd}) in the case (\ref{118ccc}).
\end{lem}
\subsection{Problem B, anti solitons for $g>g_1$.}
Assumption (\ref{20193}) guarantees the existence of a real root $g_1$ for any value of the constant $C$. We set
\be
\frac{d g}{d\eta}=-\sqrt{F(g,q)},\quad \eta\in(0,\infty);\quad
 g|_{\eta=0}=g_1.\label{2018b}
 \ee
Since $g_1>1$, instead of (\ref{118}) we obtain now the condition
 \be
 p_1\ipd 1-g_1^r=\frac{c_3A}{\gamma+\alpha^2V}<0.\label{1181}
\ee
Consequently,  (\ref{1181}) requires the restriction
\be
A<0.\label{118k}
\ee
Let us assume the fulfilment the conditions (\ref{5}), (\ref{118cc}). Then we obtain the formula for the wave speed
\be
V=\frac{1}{\alpha^2}\Big\{\frac{c_3}{p_1}A-\gamma\Big\},\label{118n}
\ee
and the counting formula for the root $g_1$
\be
F(g_1,q(g_1,A))=0\; \text{with}\; g_1>1\;\text{and}\;
 q(g_1,A)=\frac{p_1}{c_1\alpha^2A^2}\big\{c_3A-p_1\gamma_\alpha\big\}.\label{118l}
\ee
In turn, condition (\ref{20193}) reinforces constraint (\ref{118k}) and entails the assumption
\be
A<A_1^*,\quad\text{where}\quad A_1^*=p_1\gamma_\alpha/c_3<0.\label{118p}
\ee
Next, in the case $\alpha=0$, the amplitude $A$ again determines the root $g_1=\bar{g_1}(A)>1$ of $F$ by the formula similar to (\ref{118cccc}). Thus, as in the previous case, we should look for the coefficient $q=q(\bar{g_1}(A))$ such that
$F(\bar{g_1}(A),q)=0$. In turn, for negative amplitudes  $q(\bar{g_1}(A))>0$. Therefore, we uniquely determine  the wave profile and  the velocity
\be
V=c_0+\frac{c_1\gamma^2}{c_3^2}q\big(\bar{g_1}(A)\big).\label{118ddd}
\ee
It remains to cheque properties (\ref{206}), (\ref{207}). Similarly to  (\ref{118a}) we obtain
\be
\om|_{\eta=0}=1,\quad \om'|_{\eta=0}=0, \quad \om''|_{\eta=0}=-\frac{r}{2p_1}g^{r-1}F'(g,q)|_{g=g_1}<0.\label{118m}
\ee
Thus, the function $\om(\eta)$ also admits the smooth even continuation on the negative half-axis. Next,  for $g=1+z$ and $0<z<<1$ the equation (\ref{2018b}) implies again the relation $\om(\eta)\sim \exp(-\sqrt{rq}\eta)$.

We come to the following analogue of Lemma 1
\begin{lem}
Under the assumptions (\ref{5}), (\ref{118cc}) we assume the fulfilment of the restriction (\ref{118p}), and in the case (\ref{5}), (\ref{118ccc}) we assume (\ref{118k}).
  Then the Cauchy problem (\ref{2018b})  determines  the soliton solution (\ref{205}) with the velocity $V=V(A)>c_0$ defined in (\ref{118n}) and in (\ref{118ddd}) respectively.
\end{lem}
\section{Non smooth solitary waves}
In order to consider non smooth waves, let us firstly transform the original equation (\ref{1}) to the divergent form
\begin{align}
&\frac{\pa }{\pa t}\left\{u-\alpha^2\ve^2\frac{\pa^2 u}{\pa x^2}\right\}\label{1110}\\
&+\frac{\pa}{\pa x}\left\{c_0u+c_1u^3
-(c_2-c_3)\Big(\ve\frac{\pa u}{\pa x}\Big)^2+\ve^2\frac{\pa^2}{\pa x^2}\Big(\gamma u-\frac{c_3}{2}u^2\Big)\right\}=0, \notag
\end{align}
 that does not require $uu_{xx}\in \mathfrak{D}'(\mathbb{R}^1)$. We use the ansatz (\ref{205})-(\ref{207}) and notation (\ref{205a}), (\ref{2012}) again and pass to  the following version of the inverse scattering problem (\ref{209}):

\textit{Determine the velocity} $V$ \textit{so that for any test function} $\vp$ \textit{the equation}
\be
(W-\frac12W^2,\vp''')-(r(q W-W^3)+\frac{c_2-c_3}{c_3}(W')^2,\vp')=0\label{209a}
\ee
\textit{admits a nontrivial continuous solution with the properties} (\ref{206}) and (\ref{207}).
\subsection{Peakons}
Peakons, that is, continuous solitary waves  with discontinuous, but bounded first derivative, belong to "regular distributions" \cite{GelfShil}. To construct such solution of (\ref{1110}) let us define the notation
\begin{equation} \label{6}
[f]=f_+(\eta)-f_-(\eta), \quad [f]|_0=f_+(\eta)|_{\eta\to+0}-f_-(\eta)|_{\eta\to-0},
\end{equation}
 for arbitrary functions $f_\pm(\eta)$. Next we write the ansatz
\begin{equation}\label{7}
u(x,t,\ve)=A\{\om_-(\eta)+[\om]H(x-Vt)\}|_{\eta=\beta(x-Vt)/\ve},
\end{equation}
where  $\om_\pm=\om_\pm(\eta)\in C^1({\mathbb{R}}_\pm^1)$ are  functions  such that:
\be
\om_\pm|_{\eta=\pm0}=1,\quad \om_\pm(\eta)\to0\quad as\quad \eta\to\pm\infty.\label{8}
\ee
We assume also that the functions $\om_\pm$ are extended on ${\mathbb{R}}^1_\mp$ in a smooth manner. Similarly to Section 2, the amplitude $A$ here is  a free parameter, and the velocity
 $V=V(A)$ should be determined.
Obviously, (\ref{8}) implies that $[\om]|_0=0$, however, to obtain a peakon we should suppose
\begin{equation}\label{11}
[\om']|_{\eta=0}\neq 0.
\end{equation}
Note now that $H^k=H$, $k\geq1$, thus
\begin{equation}\label{12}
u^k(x,t,\ve)=A^k\{\om_-^k(\eta)+[\om^k]H(x-Vt)\}|_{\eta=\beta(x-Vt)/\ve}.
\end{equation}
Let us define $W_\pm=p\om_\pm$ with $p$ described in (\ref{2012}), and recall  how to calculate the weak derivative for a function of the form (\ref{7}): for any $\vp(\eta)\in C_0^\infty$
\begin{align}
&\Big(\frac{\pa u}{\pa \eta},\vp(\eta)\Big)\ipd -\Big(u,\frac{\pa \vp(\eta)}{\pa \eta}\Big)=-\frac Ap\int_{-\infty}^{0}W_{-}(\eta)\frac{\pa \vp(\eta)}{\pa \eta}d\eta\notag\\
&-\frac Ap\int^{\infty}_{0}W_{+}(\eta)\frac{\pa \vp(\eta)}{\pa \eta}d\eta
=\frac Ap\big(W_{+}(\eta)|_{\eta=+0}-W_{-}(\eta)|_{\eta=-0}\big)\vp(0)\notag\\
&+\frac Ap\int_{-\infty}^{0}\frac{\pa W_{-}(\eta)}{\pa \eta}\vp(\eta)d\eta
+\frac Ap\int^{\infty}_{0}\frac{\pa W_{+}(\eta)}{\pa \eta}\vp(\eta)d\eta\notag\\
&=A[\om]|_{\eta=0}\big(\delta(\eta),\vp(\eta)\big)+\frac Ap\Big(\frac{\pa W_{-}}{\pa \eta}+\Big[\frac{\pa W}{\pa \eta}\Big]H(\eta),\vp(\eta)\Big).\notag
\end{align}
Calculating next all the terms in  (\ref{209a}), we obtain a linear combination of $H(\eta)$, $1-H(\eta)$, $\delta(\eta)$, and $\delta'(\eta)$ functions.
Then the result of substitution of (\ref{7}) into (\ref{1110}) can be easily transformed to the following form:
\begin{align}
\big\{\mathfrak{W}_-+[\mathfrak{W}]H\big\}&+\big\{ [W']|_0-\frac12[(W^2)']|_0\big\}\delta' \notag\\
&+\big\{ [W'']|_0-\frac{c_2-c_3}{c_3}[(W')^2]|_0-\frac12[(W^2)'']|_0\big\}\delta=0,\label{16}
\end{align}
where
\begin{equation}
\mathfrak{W}\pm=\frac{d}{d\eta}\Big\{rW_\pm^3 -rqW_\pm+W_\pm''-\frac{c_2-c_3}{c_3}(W_\pm')^2
-\frac12\big(W_\pm^2\big)''\Big\}.\label{17}
\end{equation}
Recall that the distributions $H$, $\delta$, and $\delta'$ are linearly independent. Thus, by virtue of  (\ref{2012}), (\ref{8}), and (\ref{16}) we deduce that:
\begin{equation}\label{18}
(1-W|_0)[W']|_0=0, \quad (1-W|_0)[W'']|_0-\frac{c_2}{c_3}[(W')^2]|_0=0.
\end{equation}
Clearly, for peakons we conclude:
\be
p=1, \quad W'_-(0)=-W'_+(0).\label{19}
\ee
Consequently, (\ref{16}) - (\ref{19}) imply the equations $\mathfrak{W}\pm=0$ for the functions $W_\pm$. Furthermore, setting $W_\pm=1-g_\pm^r$ and analyzing the equation of the form  (\ref{2018aa}), we obtain the condition $C\le0$. Let
\be
C=0.\label{20195}
\ee
Then  $\om_\pm|_{\eta\to\pm 0}\to 1$, however the first derivative is not continuous,
$$
\om_\pm'|_{\eta\to\pm 0}\to\mp rg_\pm^{r-1}\sqrt{F}|_{g\to\pm0}=\mp r\sqrt{(1-q)/(1-r)}.
$$
On the contrary, if
\be
C<0\;\text{and}\;p=1,\;\text{ then}\; \sqrt{F}|_{g=0}\neq0\;\text{and}\;\om'|_{\eta\to\pm 0}\to\mp \infty. \label{21a}
\ee
Let $\alpha>0$. Then the condition (\ref{20195}) and the second equality in (\ref{2012}) allow us to determine the wave velocity
\begin{equation}
V=\frac{1}{\alpha^2}(c_3A-\gamma),\label{21}
\end{equation}
and the right-hand side of the equation (\ref{2018})
\begin{equation}
F=\frac{2}{2+r}g^{2-2r}(1-g^r)^2\big(1+\frac32r-g^r\big).\label{21b}
\end{equation}
Accordingly, we obtain the desired problem for the  function $\om_+$
\begin{equation}
\frac{d\om_+}{d\eta}=-\zeta\om_+\sqrt{\om_++\frac{3r}{2}},\quad\eta>0,\quad\om_+|_{\eta=0}=1,\label{31}
\end{equation}
where $\zeta=r\sqrt{2/(2+r)}$. Therefore,
\begin{equation}
\om_\pm=\frac{3r}{2}\sinh^{-2}(\zeta_1\eta\pm c_0),\quad \eta\in \mathbb{R}^1_\pm,\label{31a}
\end{equation}
where $\zeta_1=\zeta\sqrt{3r/2^3}$ and the constant of integration $c_0$ is such that
$$
\sinh^{2}(c_0)=3r/2.
$$
In turn, the equalities  (\ref{2012}), (\ref{20195}), and (\ref{21})  are compatible if and only if
\begin{align}
&\text{for}\quad c_3^2>4\xi\gamma_\alpha\quad A=A_0^+\quad\text{or}\quad A=A_0^-,\label{34}\\
&\text{for}\quad c_3^2=4\xi\gamma_\alpha\quad A=\bar{A_0}^\pm,\label{34a}
\end{align}
where $A_0^\pm$ and $\bar{A_0}^\pm$ are defined in  (\ref{118h}), (\ref{118gg}) with $p_0=1$. Note that if $c_3^2<4\xi\gamma_\alpha$, then the condition $C=0$ cannot be realized.

If $\alpha=0$, then the equalities  (\ref{2012}) and (\ref{20195}) imply the restrictions
\begin{equation}
A=\frac{\gamma}{c_3}, \quad V=c_0+3c_1\frac{r^2\gamma^2}{c_3^2(2+r)}.\label{21a20}
\end{equation}
Thus, we establish
\begin{lem}
Let $\alpha>0$ and the wave amplitude satisfy the conditions (\ref{34}), (\ref{34a}).
 Then  the equation (\ref{1110})  has the peakon solution (\ref{7}), (\ref{31a}) with the velocity  (\ref{21}). If $\alpha=0$, then the peakon solution exists in the case (\ref{21a20}) only.
\end{lem}
\subsection{Cuspons}
To construct a cuspon-type traveling wave we take into account (\ref{21a}) and use the ansatz (\ref{7}) again setting
\be
\om_\pm(\eta)=W_\pm(\eta)=1-g_\pm^r(\eta), \; g_\pm(0)=0,\; g_\pm(\eta)\to1\;\text{as}\; \eta\to\pm\infty. \label{35}
\ee
For $\alpha>0$ the second assumption in (\ref{35}) and (\ref{2012}) imply  the Rankine-Hugoniot type condition (\ref{21}) for the cuspon speed. Therefore, the main question for such waves is the smoothness of $\om_\pm$ and the sense in which equation (\ref{209a}) should be understood. The smoothness of the functions $\om_\pm(\eta)$ depends on the parameter $r$. Let us consider the possible cases separately.
\subsubsection{The case $c_3>c_2$}
In view of  (\ref{205a}), (\ref{35}) we obtain that $r>1/2$ and $\om'_\pm\sim \pm |\eta|^{r-1}$ for $|\eta|<<1$. Thus, $\om'_\pm\in L^2(\mathbb{R}^1_\pm)$. Therefore, all singularities in (\ref{209a}) are regular, which allows us
to convert (\ref{209a}) into the standard for distributions like $\eta_\pm^\lambda$, $\lambda\in(0,1)$, form: for each test function $\vp=\vp(\eta)$
\begin{align}
&\lim_{\mu\to0}\big\{\int_{-\infty}^{-\mu}+\int_{\mu}^\infty\big\}\Big\{\Big(\om_\pm-\frac{1}{2}\om_\pm^2\Big)\vp'''\notag\\
&-\Big(rq\om_\pm-r\om_\pm^3+\frac{c_2-c_3}{c_3}(\om_\pm')^2\Big)\vp'\Big\}d\eta=0. \label{37}
\end{align}
Next, taking into account the  conditions (\ref{35}) we get
\begin{align}
&\big\{\int_{-\infty}^{-\mu}+\int_{\mu}^\infty\big\}\Big(\om_\pm-\frac{1}{2}\om_\pm^2\Big)\vp'''d\eta =-\Big([\om_\pm-\frac{1}{2}\om_\pm^2]\big|_\mu\Big)\vp''(0)\notag\\
&-\big\{\int_{-\infty}^{-\mu}+\int_{\mu}^\infty\big\}\Big(\om_\pm'-\om_\pm \om_\pm'\Big)\vp''d\eta\notag\\
&=[(1-\om_\pm )\om_\pm']\big|_\mu\vp'(0)
+\big\{\int_{-\infty}^{-\mu}+\int_{\mu}^\infty\big\}\Big(\om_\pm''-\big(\om_\pm \om_\pm'\big)'\Big)\vp'd\eta. \label{37a}
\end{align}
Thus, the equality (\ref{12}) can be converted to the  form
\be
\lim_{\mu\to0}\big\{\int_{-\infty}^{-\mu}+\int_{\mu}^\infty\big\}\mathfrak{W}_\pm\vp'd\eta=0, \label{37b}
\ee
where $\mathfrak{W}_\pm$ is defined in (\ref{17}). It is clear that (\ref{37b}) implies the equations of the form (\ref{2013}) for the functions $W_\pm(\eta)$, $\eta\in\mathbb{R}_{\pm}^1$. Consequently, we get again the problem (\ref{2018aa}), (\ref{2018a}) with $g_0=0$. It is easy to establish now that for $|\eta|<<1$
\begin{equation}\label{37c}
g_\pm(\eta)=\sqrt{|C|}|\eta|+O(|\eta|^{3-2r}),\; \om_\pm'(\eta)=\mp r|C|^{r/2}|\eta|^{r-1}+O(|\eta|^{1-r}).
\end{equation}
Thus, if $r\in(1/2,1)$, then $g_\pm\in C^1(\mathbb{R}_{\pm}^1)$ and $\om_\pm'\in L^2(\mathbb{R}_\pm^1)$.
\subsubsection{The case $c_3<c_2$}
Now $r\in(0,1/2)$ and $\om''_\pm\sim \pm |\eta|^{r-2}$ is derived from the regular distribution $\om'_\pm\sim \pm |\eta|^{r-1}$ but it is not a regular distribution in itself. Let us recall the standard definition \cite{GelfShil} of  functions $\eta_+^{\lambda-1}=\eta^{\lambda-1} H(\eta)$ with $\lambda\in(-1,0)$
\begin{align}
&((\eta_+^\lambda)',\vp(\eta))\ipd-(\eta_+^\lambda,\vp'(\eta))=-(\eta_+^\lambda,(\vp(\eta)-\vp(0))')\notag\\
&=-\lim_{\mu\to0}\int_\mu^\infty \eta^\lambda(\vp(\eta)-\vp(0))'d\eta\label{3337c}\\
&=-\lim_{\mu\to0}\Big\{\eta^\lambda(\vp(\eta)-\vp(0))|_{\mu}^\infty-\int_\mu^\infty \big(\eta^\lambda)'(\vp(\eta)-\vp(0))\big)d\eta\Big\}\notag\\
&=\int_0^\infty \big(\eta^\lambda)'(\vp(\eta)-\vp(0))\big)d\eta=\lambda(\eta_+^{\lambda-1},\vp(\eta)-\vp(0)).\notag
\end{align}
Similarly (\ref{3337c}) we define
\begin{align}
&(\om_+''',\vp(\eta))=-(\om_+,\vp'''(\eta))=\lim_{\mu\to0}\int_\mu^\infty \om_+'\big(\vp'(\eta)-\vp'(0)\big)'d\eta\notag\\
&=\lim_{\mu\to0}\Big\{ \om_+'(\vp'(\eta)-\vp'(0))|_{\mu}^\infty-\int_\mu^\infty \om_+''\big(\vp'(\eta)-\vp'(0)\big)d\eta\Big\}\label{37d}\\
&=-\big(\om_+'',(\vp'(\eta)-\vp'(0))\big).\notag
\end{align}
Furthermore, for $\Psi_+(\mu,\eta)=\int_\mu^\eta(\om_+'(z))^2dz$ we get
\begin{align}
&\big((\Psi_+(0,\eta))'',\vp(\eta)\big)=(\Psi_+(0,\eta),\vp''(\eta))\notag\\
&=\lim_{\mu,\mu_1\to0}\int_\mu^\infty \Psi_+(\mu_1,\eta)\big(\vp'(\eta)-\vp'(0)\big)'d\eta\label{37e}\\
&=-\lim_{\mu\to0}\int_\mu^\infty \big(\om_+'(\eta)\big)^2\big(\vp'(\eta)-\vp'(0)\big)d\eta=-\big((\om_+'(\eta))^2,(\vp'(\eta)-\vp'(0))\big).\notag
\end{align}
\subsubsection{The case $c_3=c_2$}
 When $c_2=c_3$, the term $(u_x')^2$ disappears from the equation (\ref{1110}). Thus, in contrast to the general case,  it is sufficient to assume only that
\be
u\in\mathfrak{D}'(\mathbb{R}^1),\quad u^3\in\mathfrak{D}'(\mathbb{R}^1).\label{33b}
\ee
Obviously, this restriction permits not only cuspons, but also much more singular solutions. In particular, the DP equation have discontinuous solutions, the so called shockpeakons \cite{Lund}.

Finally let us note that
\be
q>3r^2/(2+r)>0 \label{333b}
\ee
 for $C<0$. Thus we obtain only the following existence condition for cuspons in the case $\alpha>0$
\begin{equation}
c_3^2>4\xi\gamma_\alpha,\quad A\in(A_0^-,A_0^+),\label{33c}
\end{equation}
where $A_0^\pm$  are defined in  (\ref{118h}) with $p_0=1$. At the same time,  cuspons do not exist in the case  $c_3^2\leq4\xi\gamma_\alpha$.

If $\alpha=0$, then the equalities  (\ref{2012}) and (\ref{21a}) imply the restrictions
\begin{equation}
A=\frac{\gamma}{c_3}, \quad V=c_0+\frac{c_1\gamma^2}{c_3^2}q.\label{21a21}
\end{equation}
Therefore, in this case cuspon can only have a fixed amplitude, whereas the parameters $q$ and $V$ are related only by the second equality (\ref{21a21}) under the condition (\ref{333b}).
This means a rather unusual situation of the existence of a family of waves with the same amplitude, but moving at different speeds and having different shapes. Indeed, by choosing any value of the parameter $q$ and calculating the function $g(\eta,q)$ in accordance with the problem (\ref{2018aa}), (\ref{2018a}) for $g_0=0$, we set the wave profile in the initial condition for the equation (\ref{1110}). According to (\ref{21a21}), the cuspon will move at a speed of $V(q)$.
\begin{figure}[H]
\centering
\includegraphics[width=13cm]{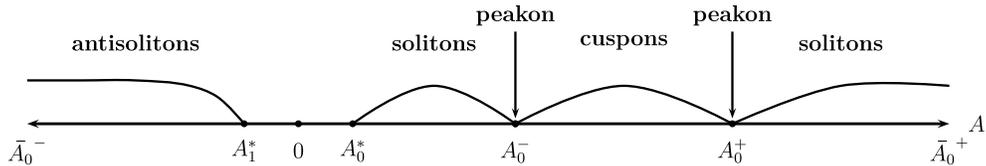}
\caption{Wave type dependent on the amplitude in the case   $\alpha>0$, $c_3^2>4\xi\gamma_\alpha$.}
\label{f2}
\end{figure}
By combining all related to cuspon, we get the statement
\begin{lem}
Assume that condition (\ref{33c}) is fulfilled for $\alpha>0$ or condition (\ref{21a21}) for $\alpha=0$. Then
 the equation (\ref{1110})  has the cuspon solution (\ref{7}), (\ref{31a}) moving at the speed of  $V=V(A)$ (\ref{21a}) or $V=V(q)$ (\ref{21a21}) respectively.
\end{lem}

\section{Conclusion}
 Let us summarize all the results of the previous sections.
 \begin{teo}
Let the conditions specified in one of Lemmas 1 - 4 be fulfilled. Then the equation (gmKdV) (\ref{1}) admits a soliton, peakon, or cuspon  solution, respectively.
 \end{teo}
  As  examples, we  present the diagrams in Fig.2 and Fig.3, which illustrate the existence of different  types of traveling waves solutions in dependence on the wave amplitude in the case $\alpha>0$, $c_3^2\geq4\xi\gamma_\alpha$. For $\alpha>0$, $c_3^2<4\xi\gamma_\alpha$ solitons exist for any amplitude $A\in \mathbb{R}^1\setminus[A_1^*,A_0^*]$, whereas non smooth waves don't exist in this case.
\begin{figure}[H]
\centering
\includegraphics[width=13cm]{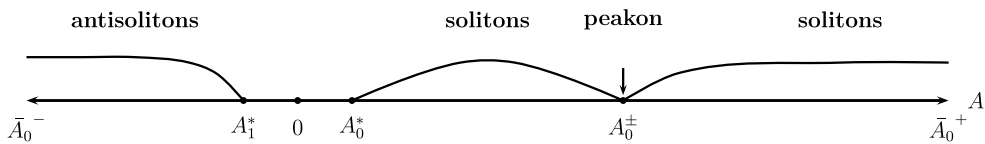}
\caption{Wave type dependent on the amplitude in the case   $\alpha>0$, $c_3^2=4\xi\gamma_\alpha$.}
\label{f3}
\end{figure}
Finally, let us list open problems for the modified Korteweg-de Vries equation:

1. Existence and uniqueness theorems for the corresponding Cauchy problem.

2. Integrability of  gmKdV equation, at least for some special items of this family.

3. Scenario of solitary wave collisions.

4. Existence of other types of traveling wave solutions.


\begin{thebibliography}{99}

\bibitem{GN}
A.~Green, P.~Naghdi,
 ``A Derivation of Equations for Wave Propagation in Water of Variable Depth",
Journal of Fluid Mechanics, {\bf 78}, 237-246 (1976).




%\bibitem{PS}
%P.~Popivanov, A.~Slavova,
% ``Peakons, cuspons, compactons, solitons, kinks and periodic solutions of several third order nonlinear PDE and their cellular neural network realization",
%Functional Differential Equations, {\bf 16} (4), 609-626 (2009).

%\bibitem{BFOD}
%J.~Bogning,  G.~Fautso Gaetan,  H.~Omanda, C.~Djeumen-Tchaho,
% ``Combined Peakons and Multiple-Peak Solutions of the Camassa- Holm and Modified KdV Equations and Their Conditions of Obtention",
%Physics Journal, {\bf 1} (3), 367-374 (2015).



\bibitem{CH}
R.~Camassa, D.~Holm,
 ``An integrable shallow water equation with peaked solitons",
 PHYS REV LETT {\bf 71}, 1661-1664 (1993).

\bibitem{CH1}
R.~Camassa, D.~Holm, and J.~Hyman,
 ``A new integrable shallow water equation",
 Adv. Appl. Mech.  {\bf 31}, 1-33 (1994).

\bibitem{GS}
O.~Glass, F.~Sueur,
 ``Smoothness of the flow map for low-regularity solutions of the Camassa-Holm equations",
 Discrete and Continuous Dynamical Systems {\bf 33} (7), 2791-2808 (2013).

\bibitem{Oct}
M.~Octavian,
 ``Existence and uniqueness of low Regularity solutions for the Dullin-Gottwald-Holm  equation",
COMMUN.MATH. PHYS. {\bf 265}, 189-200 (2006).


\bibitem{BSSz}
R.~Beals,  D.~Sattinger, J.~Szmigielski,
 ``Multipeakons and the classical moment problem",
Adv. Math.  {\bf 154}, 229-3257 (2001).

\bibitem{ConLan}
A.~Constantin, D.~Lannes,
  ``The hydrodynamical relevans of the Camassa-Holm and Degasperis-Procesi equations",
ARCH RATION MECH AN {\bf 192}, 165-186 (2009).




\bibitem{DegProc}
A.~Degasperis,  M.~Procesi,
 ``Asymptotic integrqability", in: Degasperis A., Gaeta G. (Eds.), {\it Symmetry and Perturbation Theory}, (Singapore, World Sientific, 23-37 1999).

\bibitem{DHH}
A.~Degasperis, D.~Holm, and  A.~Hone,
 ``A new integrable equation with peakon solutions'',
 Theoretical and Mathematical Physics {\bf 133}, 14631474 (2002).




\bibitem{Lund}
H.~Lundmark,
  ``Formation and dynamics of shock waves in the Degasperis-Procesi equation",
  Journal of Nonlinear Science {\bf 17} (3), 169-198 (2007).



\bibitem{ELY}
J.~Esher, Y.~Liu, Z.~Yin,
 ``Global weak solutions and blow-up structure for the Degasperis-Procesi equation",
J FUNCT ANAL {\bf 241} (2), 457-485 (2006).


\bibitem{NoyOm}
J.~Noyola Rodriguez, G.~Omel'yanov,
  ``General Degasperis-Procesi equation and its solitary wave solutions", CHAOS SOLITON FRACT {\bf 118}, 41-46 (2019).


\bibitem{Om}
 G.~Omel'yanov,
  ``Classical and Nonclassical Solitary Waves in the General Degasperis-Procesi Model",
  Russian Journal of Mathematical Physics, {\bf 26} (3), 384-390, (2019)


\bibitem{Om1}
G.~Omel'yanov, ``Collision of solitons in non-integrable versions of the Degasperis-Procesi model",
Chaos, Solitons and Fractals,  {\bf 136}, 109802, (2020)

\bibitem{NoyOm1}
J.~Noyola Rodriguez, G.~Omel'yanov,
  ``A finite difference scheme for smooth solutions of the general Degasperis-Procesi equation", Numerical Methods for Partial Differential Equations, {\bf 36} (4), 887-905 (2020).




\bibitem{Len}
J.~Lenells,
 ``Traveling wave solutions of the Camassa-Holm
equation", J. Differential Equations {\bf 217}, 393-430 (2005).

\bibitem{GH}
K.~Grunert, H.~Holden,
  ``The general peakon-antipeakon solution for the Camassa-Holm equation",
Journal of Hyperbolic Differential Equations, {\bf 13} (2), 353-380 (2016).


\bibitem{Mat}
Y.~Matsuno,
  ``Multisoliton solutions of the Degasperi-Procesi equation and their peakon limit",
RES MEAS AP {\bf 21}, 1553-1570 (2005).


\bibitem{ZQQ}
Zh.~Qiao,
 ``M-shape peakons, dehisced solitons, cuspons and new 1-peak solitons for the Degasperis-Procesi equation",
  CHAOS SOLITON FRACT {\bf 37} (2), 501-507 (2008).





\bibitem{GelfShil}
I.~M.~Gel'fand, G.~E.~Shilov,
  {\it Generalized functions}  (Academic Press, NY, 1964).



\end{thebibliography}
\end{document}